\documentclass[preprint,showpacs]{revtex4}
\usepackage{graphicx}
\usepackage{amsmath}
\usepackage{amssymb}
\usepackage{bm}
\usepackage{color}
\def\l{\langle}
\def\r{\rangle}

\begin{document}
\title{Phase transition of two-dimensional generalized XY model} 
\author{Yukihiro Komura}
\email{y-komura@phys.se.tmu.ac.jp}
\author{Yutaka Okabe}
\email{okabe@phys.se.tmu.ac.jp}
\affiliation{Department of Physics,
 Tokyo Metropolitan University, Hachioji, Tokyo 192-0397, Japan }

\date{\today}

\begin{abstract}
We study the two-dimensional generalized XY model 
that depends on an integer $q$ by the Monte Carlo method. 
This model was recently proposed by Romano and Zagrebnov. 
We find a single Kosterlitz-Thouless (KT) transition 
for all values of $q$, in contrast with the previous speculation 
that there may be two transitions, one a regular KT transition 
and another a first-order transition at a higher temperature.
We show the universality of the KT transitions by comparing 
the universal finite-size scaling behaviors 
at different values of $q$ without assuming a specific universal form 
in terms of the KT transition temperature $T_{\rm KT}$. 
\end{abstract}


\maketitle

\section{Introduction}

A unique phase transition, the Kosterlitz-Thouless (KT) 
transition \cite{Bere,KT,KT2}, occurs for the two-dimensional (2D) 
spin system, when the relevant number of spin components is two. 
It does not have a true long-range order due to the Mermin Wagner 
theorem, but the correlation function decays as a power of the distance 
at all the temperatures below the KT transition point $T_{\rm KT}$.  
Topological defects of bounded vortex-antivortex pairs exist 
at temperature lower than $T_{\rm KT}$, whereas vortices can be 
generated freely at temperature higher than $T_{\rm KT}$.

If the spin is restricted to the $xy$ components $(S^x, S^y)$, 
the model is called the {\it planar rotator model}.  On the other hand, 
if the spin is allowed to take the $z$ direction but the pair interaction is 
still given by $-J (S_i^x S_j^x+S_i^y S_j^y)$, it is called 
the classical {\it XY model}. 
Here, $J>0$ is the ferromagnetic coupling constant, and $i, j$ 
denote the sites of the spin. 
Although the terminologies of two models are sometimes used loosely, 
we distinguish them here.  
Precise Monte Carlo studies were reported for the 2D planar rotator 
model \cite{Weber,Olsson,Hasen97,Hasen05}.

In 1984, Domany {\it et al.} \cite{Domany} proposed a generalization of 
the planar rotator model; the Hamiltonian of this model is given by
\begin{equation}
  H = 2J \sum_{\l i,j \r} \Big[ 1-\big( \cos^2(\frac{\phi_i-\phi_j}{2}) 
      \big)^{p^2} \Big] ,
\end{equation}
where $\l i,j \r$ denotes the nearest neighbor pair. For $p^2=1$ the model reduces to 
the usual planar rotator model. 
When $p^2$ is larger, the potential becomes sharper. 
Using the Monte Carlo simulation they showed that the phase transition 
changes from the KT transition to the first-order transition for $p \ge 10$. 
Quite recently, the model by Domany {\it et al.} was 
revisited \cite{Sinha2010a,Sinha2010b}.

Romano and Zagrebnov \cite{Romano} proposed a generalization of 
the XY model, the {\it generalized XY model}, recently. 
They introduced an integer parameter $q>0$, and the Hamiltonian is given by 
\begin{equation}
 H = - J \sum_{\l i,j \r} (\sin \theta_i \sin \theta_j)^q \cos (\phi_i - \phi_j),
\label{Hamiltonian}
\end{equation}
where $\phi_i$ and $\theta_i$ are the azimuthal and polar angles, respectively; 
that is, three components of a spin are represented by
\begin{equation}
 (S^x, S^y, S^z) = (\sin \theta \cos \phi, \sin \theta \sin \phi, \cos \theta).
\end{equation}
We should note that $q=1$ corresponds to the usual classical XY model. 
Moreover, the $q=0$ model is equivalent to the planar rotator model.
This generalized XY model was studied by analytical approaches 
such as the mean-field approximation \cite{Romano} 
and the self-consistent harmonic approximation \cite{Mol03}. 
Moreover, the Monte Carlo studies were made for the three-dimensional 
\cite{Chamati05} and 2D \cite{gXY_MC} generalized XY models. 
In Ref.~\cite{gXY_MC}, the vortex density, specific heat, and other 
quantities were calculated.  The KT transition in the $xy$ plane was 
obtained, and the decrease of $T_{\rm KT}$ with the increase of $q$ 
was discussed.  The authors of Ref.~\cite{gXY_MC} observed 
another transition above $T_{\rm KT}$; 
there is a possibility that this transition becomes first order 
for large $q$, but it is not conclusive.  They mentioned that 
their study was only a first step in direction to a more elaborated theory. 

There are several points to be clarified for the 2D generalized XY model.  
First, the existence of a high-temperature transition 
should be checked; if there is a high-temperature transition, 
one may ask the order of the transition, a first order or 
a second order.  
One may also ask whether some symmetry is broken 
in the high-temperature transition.  
Second, it is worth studying the role of out-of-plane fluctuations 
in the phase transition as a function of $q$.  
Third, the universality of the KT transitions for various $q$ 
is an interesting problem.  

In this paper, we study the 2D generalized XY model, 
whose Hamiltonian is given by Eq.~(\ref{Hamiltonian}), 
on the square lattice with periodic boundary conditions 
by the Monte Carlo simulations.  
% Periodic boundary conditions are applied. 
We use both the canonical Monte Carlo method 
using the cluster flip \cite{SW,Wolff} and 
the Monte Carlo method to calculate the energy density of states (DOS) 
directly, that is, the Wang-Landau method \cite{WL}. 
We carefully check the phase transition, and discuss the universality 
of the KT transition. 
The rest of the present paper is organized as follows. 
Sec. II describes the details of the Monte Carlo method. 
The result is given in Sec. III.  The last section is devoted 
to the summary and discussions. 

\section{Monte Carlo Methods}

Both the hybrid canonical Monte Carlo method and 
the Wang-Landau method \cite{WL} are employed in the present work. 
We use the terminology of "hybrid" because we combine 
the cluster flip in the $xy$ plane and the single spin flip 
for the $z$ component.  Since the correlation length becomes longer 
in the $xy$ plane, we employ the Wolff flip of the embedded cluster 
formalism \cite{Wolff} in the $xy$ plane.  To make a cluster flip, 
we select the mirror plane perpendicular to the $xy$ plane, 
and the trial spin configuration is obtained by reflecting 
all the spins on the Kasteleyn-Fortuin cluster \cite{KF}. 
We also make a Metropolis single spin flip to allow the update of 
$z$ components.  We should note that the choice of trial spins 
should be selected uniformly among the solid angle $4 \pi$ 
in the $xyz$ sphere.  This hybrid Monte Carlo method is 
essentially the same as the method used in Ref.~\cite{gXY_MC}. 

We make the hybrid Monte Carlo simulations of the generalized 
XY model for $q$ = 0, 1, 3, 5 and 10 with the linear system size 
of $L$ = 16, 32, 64 and 128.  Typical Monte Carlo steps per spin 
are the first 50,000 MCS per spin for equilibration and 
the subsequent 200,000 MCS per spin for taking thermal average. 
One MCS consists of both cluster flip and single-spin flip. 
We calculate the energy $E$, the specific heat $C$ and 
the ratio of the correlation functions with different distances. 
This correlation ratio is a good estimator for studying 
the KT transition~\cite{Tomita02}.
We made five independent runs to estimate error bars.

Since the possibility of the first-order transition was suggested,
we also use the Monte Carlo method to calculate 
the energy DOS directly.  This type of the Monte Carlo 
method is effective for the study of a first-order transition. 
The multicanonical simulation for the 2D ten-state Potts model 
\cite{Berg} correctly estimated the interfacial free energy, 
which was later proved by the explicit formula \cite{Borgs}.  
A refined Monte Carlo algorithm to calculate the DOS was proposed 
by Wang and Landau \cite{WL}; the Wang-Landau method was successfully 
applied to several problems \cite{Yamaguchi01,Okabe06}. 
In the Wang-Landau method, a random walk in energy space 
is performed with a probability proportional 
to the reciprocal of the DOS, $1/g(E)$, which results in 
a flat histogram of energy distribution.  
Since the DOS is not known {\it a priori}, it is tuned during 
the simulation with introducing a large modification factor $f$.
The modification factor is gradually reduced to unity 
by checking the `flatness' of the energy histogram. 

In the present paper, we use the Wang-Landau method, 
and investigate the order of the transition from the information of 
the energy DOS.  The systems we treat are $q$= 1, 3, 5, 10, 30, 50 
and 100 with $L$ = 16 and 32.  In some cases, we also treat 
larger sizes. A flatness condition is that 
the histogram for all possible $E$ is not less
than 80\%, and we set the final value of ($\ln f$) 
as $10^{-8}$ following the original paper by Wang and Landau \cite{WL}.
Since we treat the system with continuous energy, we set the energy bin 
for the DOS as $0.5 J$. 
The obtained energy DOS is normalized appropriately. 
We do not care about the constant shift in $\ln(g(E))$.

\section{Results}

\subsection{Energy and Specific Heat}
\begin{figure}[tb]
\begin{center}
\includegraphics[width=14.0cm]{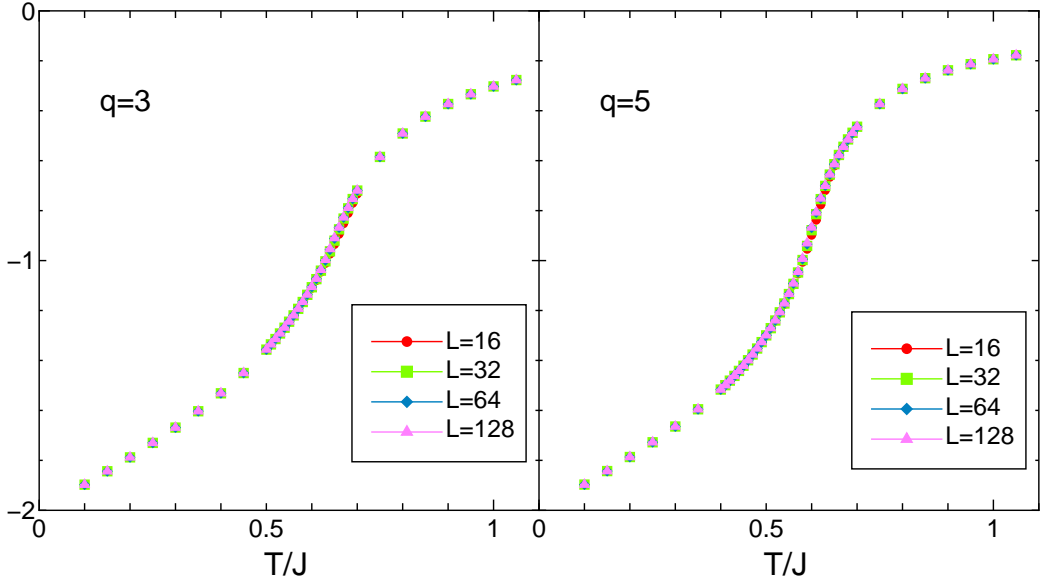}
\end{center}
\caption{Temperature dependence of the energy of 
the 2D generalized XY model for $q$=3 and $q$=5. 
The system sizes $L$ are 16, 32, 64 and 128.}
\label{fig_1}
\end{figure}

\begin{figure}[tb]
\begin{center}
\includegraphics[width=14.0cm]{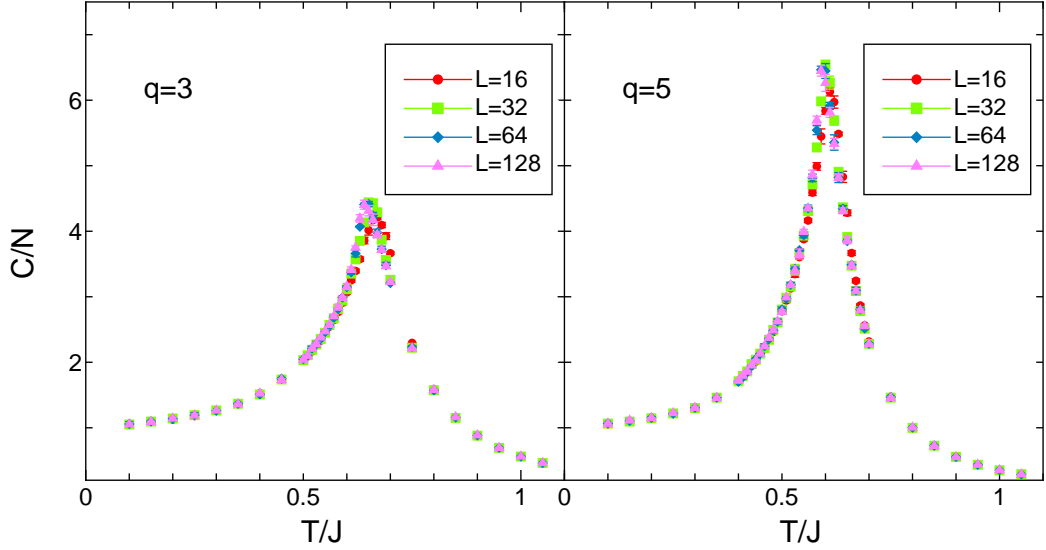}
\end{center}
\caption{Temperature dependence of the specific heat of 
the 2D generalized XY model for $q$=3 and $q$=5. 
The system sizes $L$ are 16, 32, 64 and 128.}
\label{fig_2}
\end{figure}

We plot the temperature dependence of the energy per spin 
obtained by the hybrid Monte Carlo method for $q=3$ and $q=5$ 
in Fig.~\ref{fig_1}. The system sizes 
are $L$= 16, 32, 64 and 128.  We also plot the specific heat
for $q=3$ and $q=5$ in Fig.~\ref{fig_2}. 
We plot the data of hybrid Monte Carlo simulation with one-sigma 
estimated error bars.  Most of error bars in Figs.~\ref{fig_1} and 
\ref{fig_2} are smaller than the size of the symbols. 
We see that the energy and the specific heat change smoothly 
as a function of temperature, and the size dependences are small. 
The transitions are weak for $q=3$ and $q=5$ generalized XY model; 
there is no sign of a first-order transition, for example, 
the steep jump of the energy.
It shows a typical behavior of the KT transition. 
This result is different from that of Ref.~\cite{gXY_MC}. 
Since our result is consistent with the calculation 
by the Wang-Landau method, which will be shown later, 
the previous results \cite{gXY_MC} may include some errors. 

\begin{figure}[tb]
\begin{center}
\includegraphics[width=7.0cm]{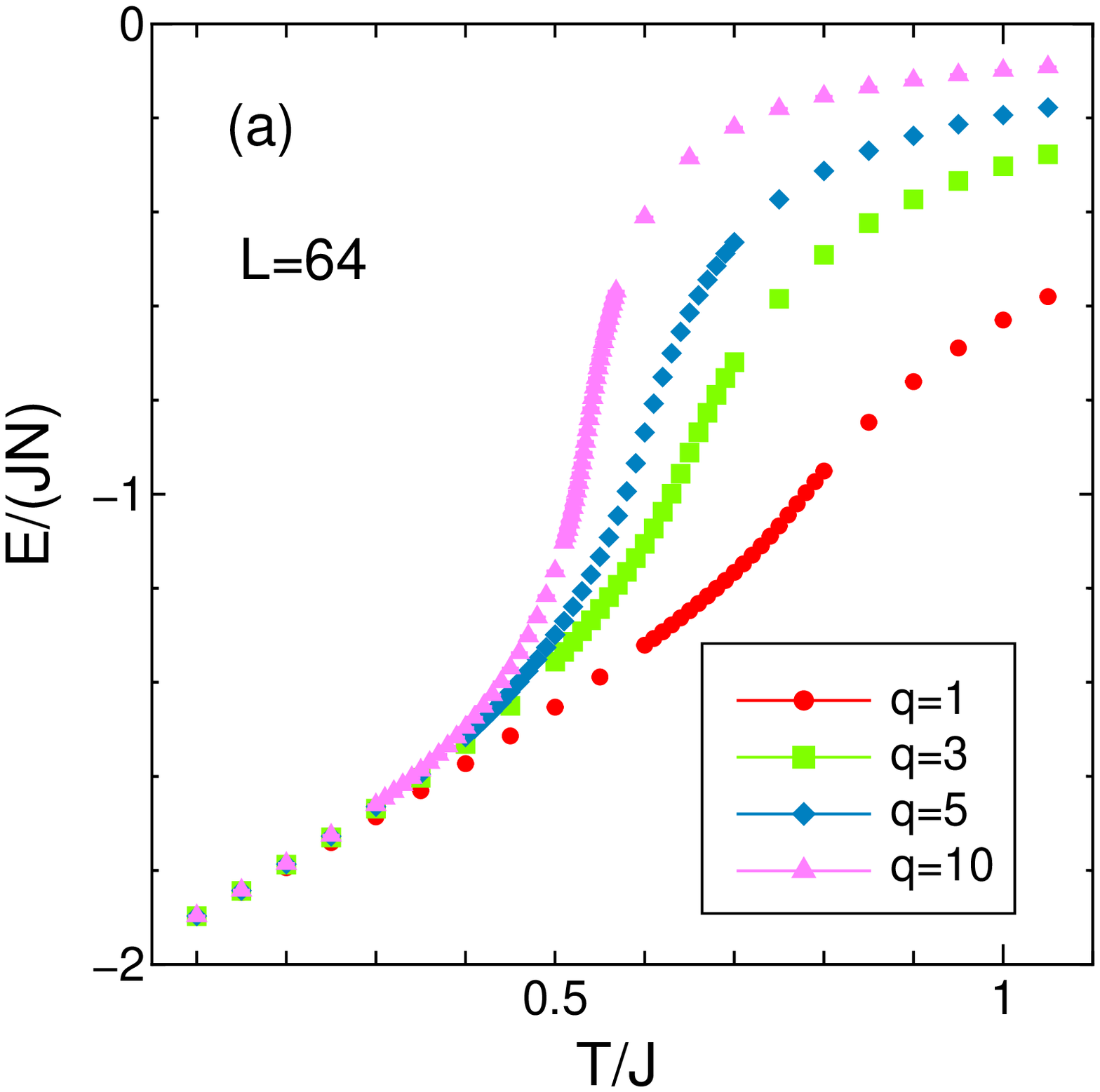}
\hspace{5mm}
\includegraphics[width=7.0cm]{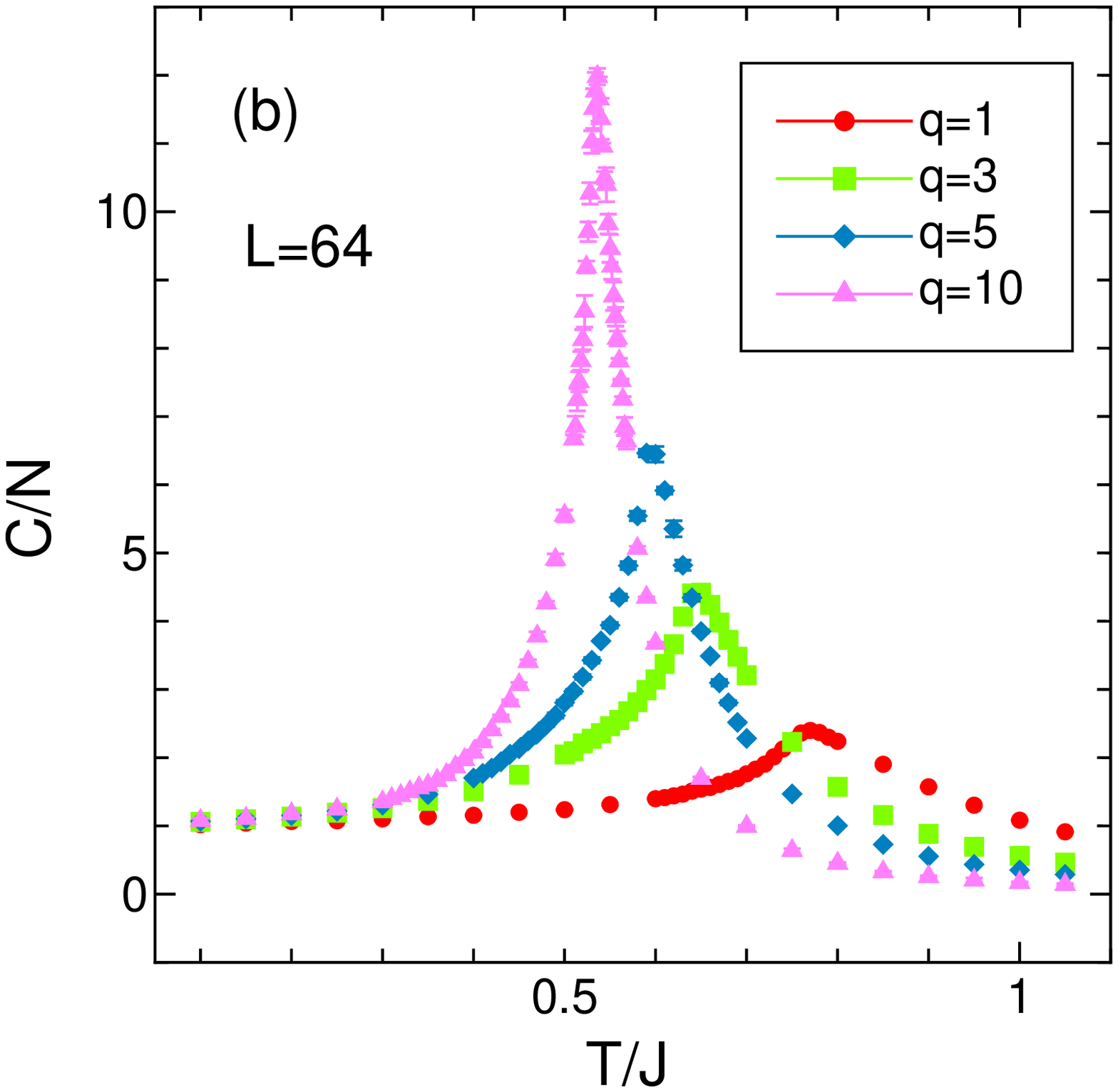}
\end{center}
\caption{Temperature dependence of the energy (a) and 
the specific heat (b) of the 2D generalized XY model 
for $L$=64 with $q$=1, 3, 5 and 10.}
\label{fig_3}
\end{figure}

To check the $q$ dependence, we plot the energy per spin 
with fixing the system size as $L=64$ for various $q$ 
in Fig.~\ref{fig_3} (a). The values of $q$ are 1, 3, 5 and 10. 
We also plot the specific heat for $q$=1, 3, 5 and 10 
in Fig.~\ref{fig_3} (b).  The system size is fixed as $L=64$ again.
We see from Fig.~\ref{fig_3} that 
the energy change becomes sharper when $q$ increases. 
The temperature which gives a peak for the specific heat 
becomes lower when $q$ increases, and the peak height becomes higher. 

\subsection{Wang-Landau method}

\begin{figure}[tb]
\begin{center}
\includegraphics[width=7.0cm]{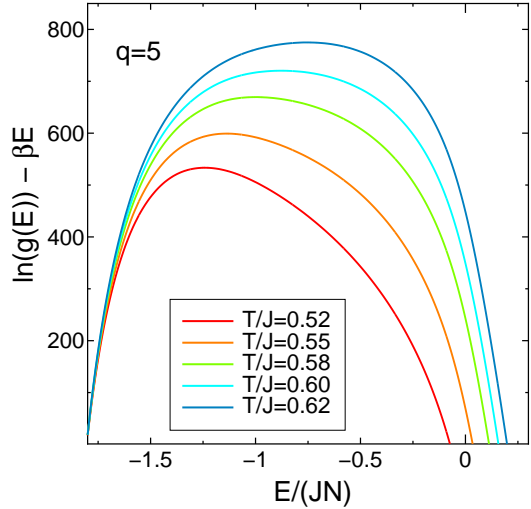}
\end{center}
\caption{Plot of $\ln(g(E)) - \beta E$ as a function of $E$ 
for $q$= 5 and $L$=32.  The data for the temperature range 
$0.52 \le T/J \le 0.62$ are plotted.}
\label{fig_4}
\end{figure}

We use the Wang-Landau method to examine the order of the transition. 
Since the authors of Ref.~\cite{gXY_MC} suggested the possibility of the 
first-order transition just above the KT transition point for $q = 5$, 
we investigate the $q = 5$ generalized XY model by using the Wang-Landau method. 
If a first-order transition occurs, the free energy, 
$F = E - (1/\beta) \ln(g(E))$, shows a double minimum 
in the thermodynamic limit at the first-order transition point $T_c$. 
Here $\beta$ is inverse temperature and $g(E)$ 
is the energy DOS, which the Wang-Landau method can directly estimate.
Since the earlier study suggested that the $q = 5$ generalized XY model 
has a first-order transition at $0.52 < T_c/J < 0.62$, 
we plot $\ln(g(E)) - \beta E$ as a function of $E$ 
in this temperature range for $L = 32$ in Fig.~\ref{fig_4}. 
We see from Fig.~\ref{fig_4} that there is no double maximum structure 
which is a sign of a first-order transition. 
For larger system sizes, the situation remains the same. 

\begin{figure}[tb]
\begin{center}
\includegraphics[width=7.0cm]{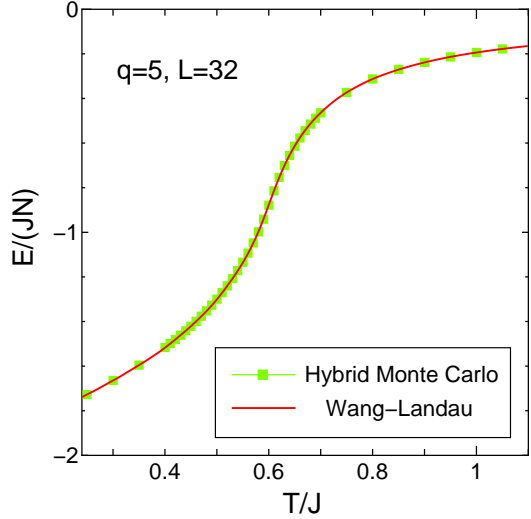}
\end{center}
\caption{Comparison of temperature dependence of $E$ for $q=5$ with $L=32$ 
calculated by the hybrid Monte Carlo method and by the Wang-Landau method.}
\label{fig_5}
\end{figure}

We compare the calculation results of energy by both hybrid Monte Carlo 
simulation and Wang-Landau method in Fig.~\ref{fig_5}, 
where $q$=5 and $L$=32. 
Both results are consistent with each other, which indicates 
the reliability of both results.

\subsection{Correlation Ratio}

Since the possibility of the extra first-order transition 
was ruled out, we next study the behavior of the KT transition. 
We here calculate 
the ratio of the correlation function \cite{Tomita02} defined as
\begin{equation}
  R(L,T) = \frac{G(L/2,T)}{G(L/4,T)},
\label{c_ratio}
\end{equation}
where $G(r,T)$ is the spin-spin correlation function 
in the $xy$ plane with the distance $r$,
\begin{equation}
  G(r,T) = \l S^{x}_{i}S^{x}_{i+r}+S^{y}_{i}S^{y}_{i+r} \r.
\end{equation}
Here, $\l \cdots \r$ denotes the thermal average.  
Starting from the scaling behavior of the correlation 
function of finite system with system size $L$,
\begin{equation}
  G(r,T) \sim r^{-(D-2+\eta)} \ h(r/L,\xi/L),
\end{equation}
we can obtain scaling properties of the correlation 
ratio $R(L,T)$, Eq.~(\ref{c_ratio}), \cite{Tomita02}.
Here, $\xi$ is the correlation length, $\eta$ is 
the decay critical exponent and $D$ is the spatial dimension. 

\begin{figure}[tb]
\begin{center}
\includegraphics[width=7.0cm]{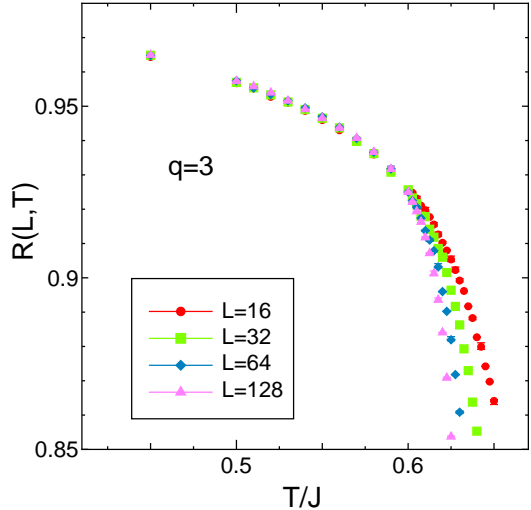}
\end{center}
\caption{Temperature dependence of the correlation ratio $R(L,T)$ 
of the 2D generalized XY model for $q$=3.
The system sizes $L$ are 16, 32, 64 and 128.}
\label{fig_6}
\end{figure}

If we plot the temperature dependence 
of the correlation ratio for the systems showing the KT transition, 
for $T<T_{\rm KT}$ the correlation ratio has no size dependence, 
but for $T>T_{\rm KT}$ it depends on $L$.  
This comes from the fact that the correlation function decays 
as a power of the distance at all the temperatures 
below the KT transition point. 
The Binder ratio \cite{Binder} has the same properties, 
but corrections are much smaller for the correlation ratio 
\cite{Tomita02}. 
Other possible choices are the helicity modulus and the ratio of 
the second moment correlation length to $L$, that is, $\xi_L/L$ \cite{Hasen05}. 
We plot the temperature dependence of the ratio of correlation function 
for $q = 3$ in Fig.~\ref{fig_6}.  The system sizes are 
$L$ = 16, 32, 64 and 128. 
We find a typical behavior of the KT transition; that is, 
for $T/J <0.6$ the correlation ratio has little size dependence, 
whereas for $T/J>0.6$ it depends on size, 
which suggests the critical temperature as $T_{\rm KT}/J \simeq 0.60$ 

\subsection{FSS of Correlation Ratio}

The finite-size scaling (FSS) is often used 
to analyze the KT transition.  Assuming the KT form of the divergence of 
the correlation length as $\xi \propto \exp(c/\sqrt{|T-T_{\rm KT}|})$, 
we may plot the correlation ratio 
as a function of $L/\exp(c/\sqrt{|T-T_{\rm KT}|})$ 
for each size $L$.  Then, if we choose $T_{\rm KT}$ (and $c$)
such that all the data are collapsed on a single curve, 
we can determine $T_{\rm KT}$.  Actually, we estimate 
$T_{\rm KT}$ as $T_{\rm KT}/J = 0.60(1)$ for $q=3$, where 
the numbers in parentheses denote the errors in 
the last digit. 

\begin{figure}[tb]
\begin{center}
\includegraphics[width=14.0cm]{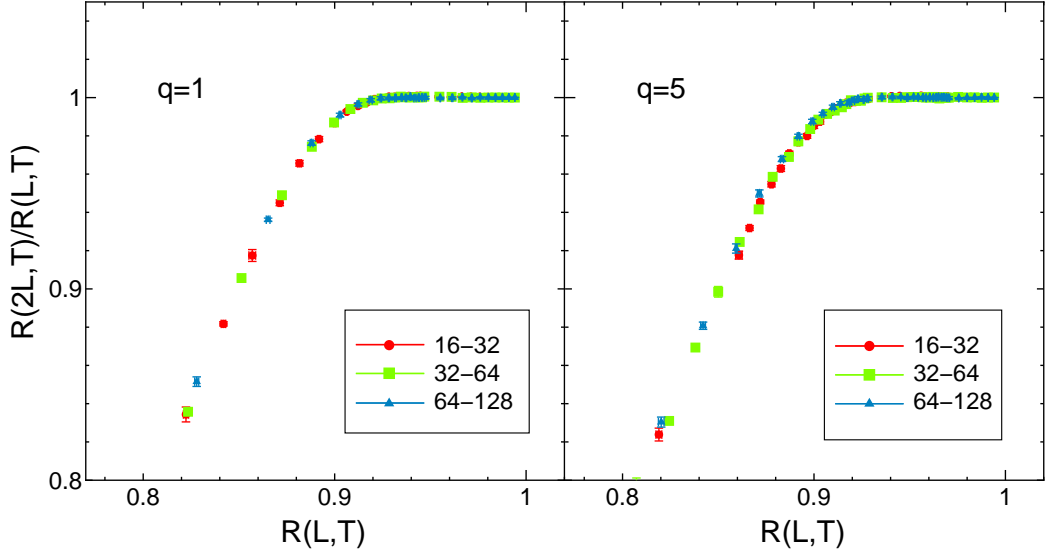}
\end{center}
\caption{FSS plot of $R(2L,T)/R(L,T)$ versus $R(L,T)$
of the 2D generalized XY model for $q$=1 and 5. 
The pair of system sizes $L$-$2L$ are 16-32, 32-64 and 64-128.} 
\label{fig_7}
\end{figure}

Here we do not go further into the usual FSS approach.  
Instead we employ the FSS without estimating $T_{\rm KT}$ 
to investigate the universality of the KT transition in detail. 
We focus on the FSS property of the ratio of the correlation ratio 
of different sizes, that is, $R(2L,T)/R(L,T)$.
Such FSS was employed in the analysis of Potts model \cite{Caraccido,Salas}, 
but no attempt has been made for the KT transition. 
We note that in the KT transition, logarithmic corrections \cite{KT2,Janke} 
might not be small. 
We plot $R(2L,T)/R(L,T)$ as a function of $R(L,T)$ in Fig.~\ref{fig_7} 
for $q=1$ and $q=5$.  
The lattice size pairs $L$-$2L$ are 
indicated in the inset of the figure, such as 16-32.  
The value of $R(2L,T)/R(L,T)$ remains 1 
for $R(L,T) \sim 1$, which corresponds to low temperature.  
The value of $R(2L,T)/R(L,T)$ deviates from 1 for smaller 
$R(L,T)$, which means $T > T_{\rm KT}$. 
All the data for $L$ = 16, 32 and 64 are collapsed 
on a single curve for both $q$=1 and $q$=5. 
That is, the FSS is very well.  

\begin{figure}[tb]
\begin{center}
\includegraphics[width=7.0cm]{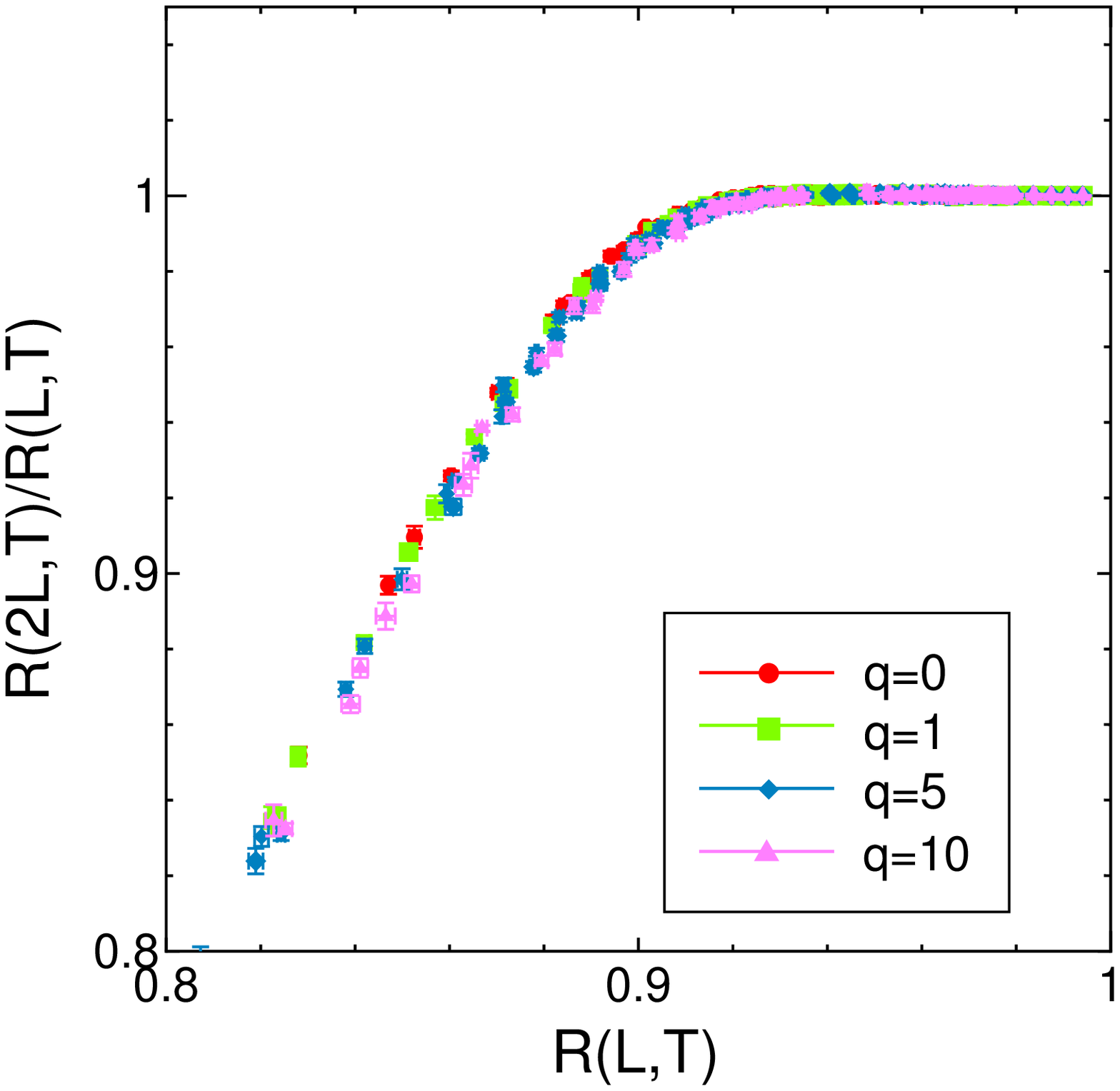}
\end{center}
\caption{Universal FSS plot of $R(2L,T)/R(L,T)$ versus $R(L,T)$
of the 2D generalized XY model for $q$=0 (planar rotator model), 
1 (XY model), 5 and 10. 
The pair of system sizes $L$-$2L$ are 16-32, 32-64 and 64-128.} 
\label{fig_8}
\end{figure}

Since the FSS plots of $q=1$ and $q=5$ look the same 
in Fig.~\ref{fig_7}, we try a universal FSS plot.
We plot both data for $q=1$ and $q=5$ in the single 
figure, Fig.~\ref{fig_8}, where system sizes are 
$L$=16, 32 and 64.  In Fig.~\ref{fig_8}, not only 
the data for $q=1$ (classical XY model) and $q=5$, 
but also those for $q=0$ (planar rotator model) and $q=10$ 
are plotted.  All the data with different $q$ are collapsed 
on a single curve within error bars.  
It means that the universal FSS is satisfied for $q$=0, 1, 5 and 10.  
This is the first direct demonstration of the universality of
the planar rotator model ($q=0$) and the classical 
XY model ($q=1$).  Thus, we conclude that the generalized XY model 
clearly shows the universality of the KT transition 
at least for $q \le 10$.  
From the universal FSS we can make the following statement. 
Although the KT transition points are different for each $q$, 
the value of the correlation ratio at the KT transition point 
is the same for all $q$.

\subsection{Large $q$ behavior}

\begin{figure}[tb]
\begin{center}
\includegraphics[width=7.0cm]{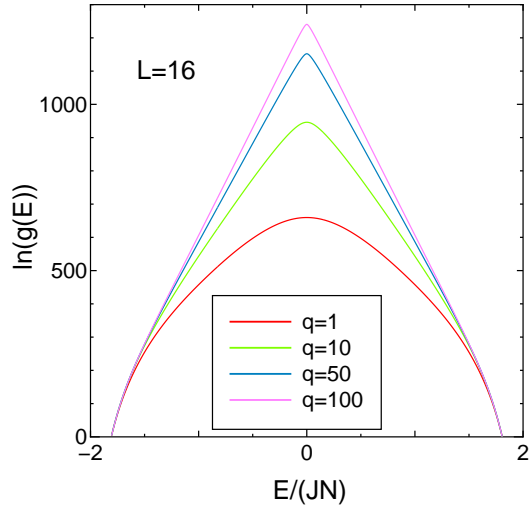}
\end{center}
\caption{Energy DOS $g(E)$ for $q$ = 1, 10, 50 and 100.
The system size $L$ is 16.} 
\label{fig_9}
\end{figure}

\begin{figure}[tb]
\begin{center}
\includegraphics[width=7.0cm]{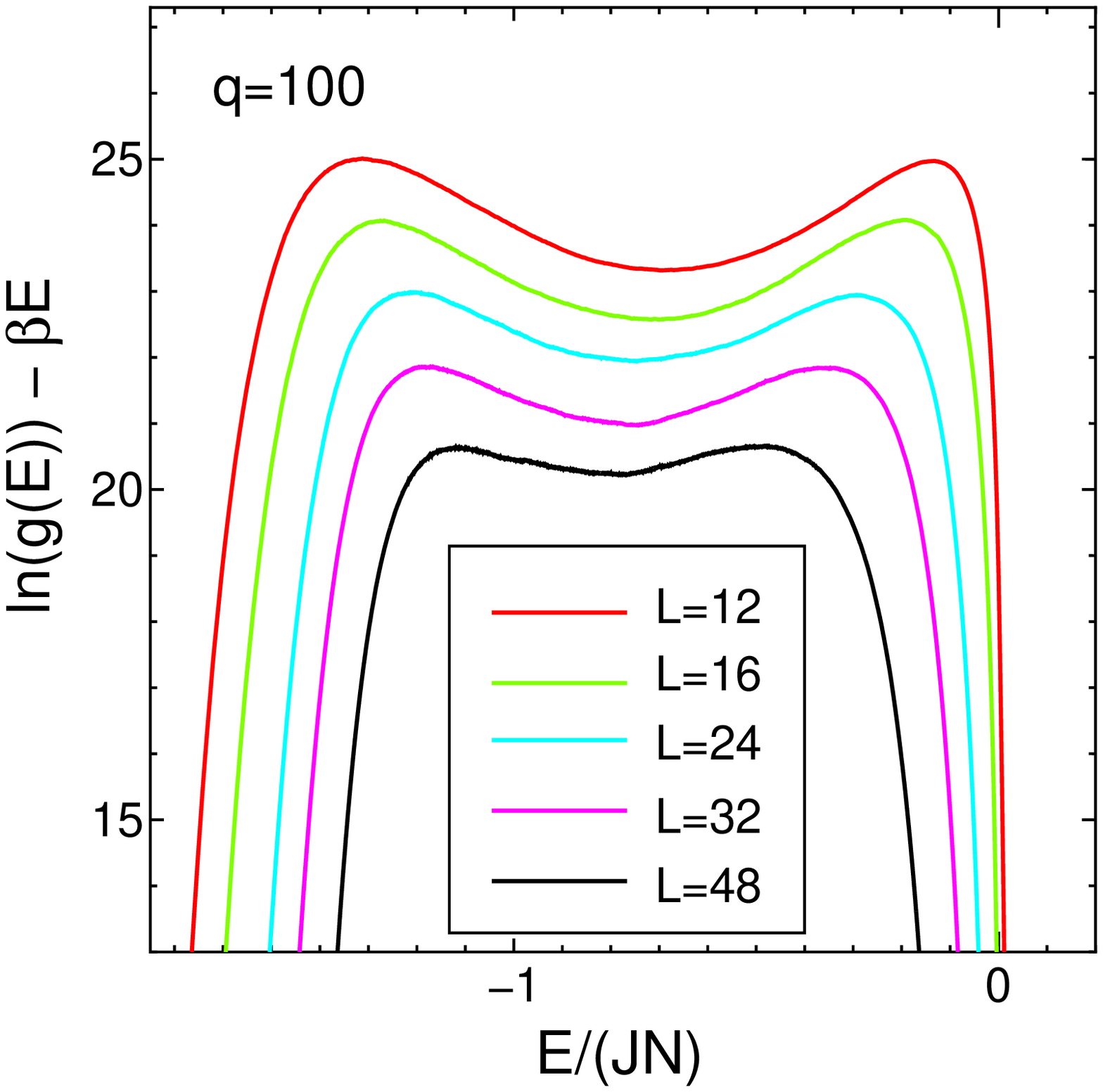}
\end{center}
\caption{
Plot of $\ln(g(E)) - \beta E$ as a function of $E$ 
for $q$= 100 and $L$=12, 16, 24, 32 and 48.  
The temperature was chosen such that the height of 
two maximum is the same.
} 
\label{fig_10}
\end{figure}

We have confirmed the universal behavior of the KT transition 
at least for $q \le 10$.  But we saw in Fig.~\ref{fig_3} 
that the transition becomes sharper for larger $q$. 
Now we study the generalized XY model for large enough $q$. 
We plot the energy DOS $\ln(g(E))$ for $q$ = 1, 10, 50 and 100 
in Fig.~\ref{fig_9}, where the system size $L$ is fixed as 16. 
We find that $\ln(g(E))$ becomes a shape of triangle for larger $q$. 
The slope of the triangle becomes steeper with the increase of $q$. 
To check whether there exists a first-order transition or not 
for $q = 100$, we follow the method shown in subsection III B. 
We give the energy dependence of $\ln(g(E))-\beta E$ 
for $L$ = 12, 16, 24, 32 and 48 in Fig.~\ref{fig_10}.  
In this plot, the data for different size are shifted vertically 
to make the structure clearly visible.  
We found a small double maximum structure in a short range 
of temperature. 
We note that the scale of vertical axis is much smaller 
than that of Fig.~\ref{fig_4} or \ref{fig_9}.
The temperature was chosen such that the height 
of two maximum is the same, 
that is, $T/J$ is 0.3957, 0.3954, 0.3949 and 0.3948 for 
$L$ = 16, 24, 32 and 48, respectively. 
Although there exists double maximum in Fig.~\ref{fig_10}, 
the difference of maximum and minimum values of 
$\ln(g(E))-\beta E$ becomes smaller with increasing the lattice size. 
If a first-order transition occurs, 
the difference of maximum and minimum values 
of $\ln(g(E))-\beta E$ is expected to be proportional to $L^{D-1}$, 
where $D$ is the spatial dimension; $D$ is two in the present study. 
This means that the $q = 100$ generalized XY model shows 
a behavior close to a first-order transition, 
but it is not a true first-order transition. 

\begin{figure}[tb]
\begin{center}
\includegraphics[width=7.0cm]{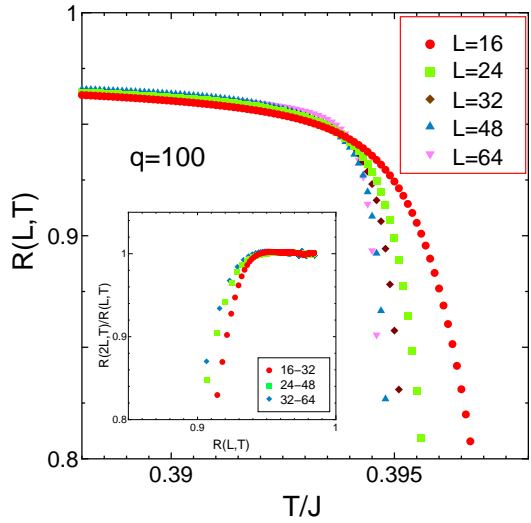}
\end{center}
\caption{Temperature dependence of the correlation ratio $R(L,T)$ 
of the 2D generalized XY model for $q$=100.
The system sizes $L$ are 16, 24, 32, 48 and 64. 
The FSS plot of $R(2L,T)/R(L,T)$ is shown in the inset.}
\label{fig_11}
\end{figure}

It is interesting to investigate the behavior of 
the KT transition for $q = 100$.  
We plot the temperature dependence of the correlation ratio 
for $L$ = 16, 24, 32, 48 and 64 in Fig.~\ref{fig_11}. 
The hybrid Monte Carlo simulation is not effective 
because the transition is close to the first order in this case. 
Thus, we have used the Wang-Landau method to calculate 
the correlation ratio in Fig.~\ref{fig_11}, although 
the system size we can treat is limited. 
Since the size dependence is small for lower temperatures, 
we can say that Fig.~\ref{fig_11} shows a behavior of KT transition, 
but we also observe a deviation from the KT behavior.  
It seems that the curves with different sizes do not merge 
but cross for low temperatures.   
The FSS plot of $R(2L,T)/R(L,T)$,
which is shown in the inset of Fig.~\ref{fig_11}, 
deviates from the universal 
FSS plot given in Fig.~\ref{fig_8}. 
Since the deviations are larger for smaller $L$, 
they can be the corrections to FSS. 
It seems that the KT transition is recovered only 
in the thermodynamic limit for large enough $q$. 

\section{Summary and Discussions}

We have studied the 2D generalized XY model on the square lattice 
using the Monte Carlo methods.  The behavior of the phase transitions 
with the change of the generalized parameter $q$ has been 
carefully checked by using both the hybrid canonical Monte Carlo 
method and the Wang-Landau method.  The results of the temperature 
dependence of the energy and the specific heat show no sign 
of the first-order transition for all $q$; moreover, the careful check 
with the Wang-Landau method yields the same result. 
We have concluded that only the KT transition occurs 
in the 2D generalized XY model, which is contrary to the conclusion 
of Ref.~\cite{gXY_MC} but is consistent with original proposal 
by Romano and Zagrebnov \cite{Romano}. 

Since it has been made clear that there exists only a single KT 
transition, we have made the FSS analysis without assuming 
the transition temperature \cite{Caraccido,Salas} 
paying a special attention to the universality.  
We have studied the ratio of the correlation ratio of two sizes; 
namely, $R(2L,T)/R(L,T)$ versus $R(L,T)$.  We have obtained that 
all the data for different $L$ are collapsed on a single curve 
for each $q$, which indicates that the FSS works well.  
Moreover, such curves with different $q$ are universal for all $q$ 
including the XY model ($q$=1) and the planar rotator model ($q$=0).  

We have made a detailed analysis for the large $q$ behavior with the 
Wang-Landau method.  When $q$ is large enough, the graph of 
the logarithm of energy DOS, $\ln(g(E))$, approaches the shape of triangle. 
Although there is a small double maximum in the plot of $\ln(g(E))-\beta E$, 
the free energy barrier becomes smaller for larger system size. 
It means that the transition is not a first order transition. 
From the analysis of the correlation ratio, the transition 
is like KT type, but the corrections to FSS are not negligible 
for larger $q$. 

From the shape of Hamiltonian, Eq.~(\ref{Hamiltonian}), 
we see that for small $q$ the microscopic states apart from 
the $xy$ plane also contribute to the $xy$ coupling 
$\cos (\phi_i - \phi_j)$ to some extent, whereas only 
the microscopic states near the $xy$ plane contribute to 
the $xy$ coupling for large $q$, due to the factor $(\sin \theta)^q$. 
This is the reason why $\ln(g(E))$ becomes a shape of triangle 
for larger $q$. In other words, for large $q$, 
the proportion of redundant states 
becomes larger as far as the $xy$ coupling is concerned. 
Recently, the role of invisible states, or the redundant states, was 
investigated for the Potts model \cite{Tamura}. 
It was discussed that the order of transition changes from 
the second order to the first order 
when the number of invisible states increases.  
It will be an interesting problem to make clear 
the role of redundant states in the case of the KT 
transition.

\section*{Acknowledgments}

We thank Toshihiko Takayama for valuable discussions. 
This work was supported by a Grant-in-Aid for Scientific Research 
from the Japan Society for the Promotion of Science.
The computation of this work has been done using computer facilities 
of Tokyo Metropolitan University and those of 
the Supercomputer Center, Institute for Solid State Physics, 
University of Tokyo.

\end{document}